\documentclass[12pt]{article}
\usepackage{amsfonts}
\makeatletter
\renewcommand{\thesection}{\Roman{section}}
\renewcommand\thesubsection{\thesection.\@arabic\c@subsection}
\newcommand{\sect}[1]{\setcounter{equation}{0}\section{#1}}

\makeatother
\def\QATOP#1#2{{#1 \atop #2}}
\def\text#1{\mbox{$#1$}}
\begin{document}
\title{\textbf{Integrability of the $C_{n}$ and $BC_{n}$ Ruijsenaars-Schneider
models}}
\author{Kai Chen\thanks{
e-mail :kai@phy.nwu.edu.cn}, Bo-yu Hou\thanks{
e-mail :byhou@phy.nwu.edu.cn} and Wen-Li Yang\thanks{
e-mail :wlyang@phy.nwu.edu.cn} \\
{\small \textit{Institute of Modern Physics, Northwest University, Xian
710069, China }}}
\date{}
\maketitle

\begin{quote}
{\bf Abstract:} We study the $C_{n}$ and $BC_{n}$ Ruijsenaars-Schneider($RS$) models with
interaction potential of trigonometric and rational types. The Lax pairs for
these models are constructed and the involutive Hamiltonians are also given.
Taking nonrelativistic limit, we also obtain the Lax pairs for the
corresponding Calogero-Moser systems.\newline
\end{quote}
\sect{Introduction}
Ruijsenaars-Schneider($RS$) and Calogero-Moser($CM$) models
as integrable many-body models recently have attracted remarkable attention
and have been extensively studied. They describe one-dimensional $n$%
-particle system with pairwise interaction. Their importance lies in various
fields ranging from lattice models in statistics physics\cite{h1, nksr}, the
field theory to gauge theory\cite{gm, n}. e.g. to the Seiberg-Witten theory
\cite{bmmm} et al. Recently, the Lax pairs for the elliptic $CM$ models in
various root system have been given by Olshanetsky et al\cite{op}, Bordner
et al\cite{bcs,bcs2,bcs3,bcs1} and D'Hoker et al\cite{hp1} respectively,
while the commutative operators for $RS$ model based on various type Lie
algebra given by Komori \cite{ko1, ko2}, Diejen\cite{di, di1} and Hasegawa%
\cite{h1, h2}et al. An interesting result is that in Ref. \cite{kai1}, the
authors show that for the $sl_{2}$ trigonometric $RS$ and $CM$ models exist the
same non-dynamical $r$ -matrix structure compared with the usual dynamical
ones. On the other hand, similar to Hasegawa's result that $A_{N-1}$ $RS$
model is related  to the $Z_n$ Sklyanin
algebra, the  integrability of	$CM$ model can be
depicted by $sl_{N}$ Gaudin algebra\cite{kai2}.

As for the $C_{n}$ type $RS$ model, commuting difference operators acting on
the space of functions on the $C_{2}$ type weight space have been
constructed by Hasegawa et al in Ref. \cite{h2}. Extending that work, the
diagonalization of elliptic difference system of that type has been studied
by Kikuchi in Ref. \cite{ki}. Despite of the fact that the Lax pairs for $CM$
models have been proposed for general Lie algebra even for all of the finite
reflection groups\cite{bcs1}, however, the Lax integrability of $RS$ model are
not clear except only for $A_{N-1}$ -type\cite{r1, nksr, bc, kz, s1, s2} and
for $C_{2}$ by the authors by straightforward construction\cite{kai3}, i.e.
the general Lax pairs for the $RS$ models other than $A_{N-1}$ -type have not
yet been obtained.

Extending the work of Ref. \cite{kai3}, the main purpose of the present paper is
to provide the Lax pairs for the $C_{n}$ and $BC_{n}$
Ruijsenaars-Schneider($RS$) models with the trigonometric and rational
interaction potentials. The key technique we used is Dirac's method
on the system imposed by some constraints. We shall give the explicit forms of
Lax pairs for these systems. It is turned out that the $C_{n}$ and $BC_{n}$
$RS$ systems can be obtain by Hamiltonian reduction of $A_{2n-1}$ and $A_{2n}$
ones. The characteristic polynomial of the Lax matrixes leads to a complete
set of involutive Hamiltonians associated with the root system of $C_{n}$
and $BC_{n}$. In particular, taking their non-relativistic limit, we shall
recover the systems of corresponding $CM$ types.

The paper is organized as follows. The basic materials about $A_{N-1}$ $RS$
model are reviewed in Sec. \ref{ars}. We also give a Lax pair associating
with Hamiltonian which has a reflection symmetry with respect to the
particles in the origin. The main results are showed in Secs. \ref{Ham} and
 \ref{lax}. In Sec. \ref{Ham}, we present the Lax pairs of $C_{n}$ and
$BC_{n}$ $RS$ models by reducing from that of $A_{N-1}$ $RS$ model. The
explicit forms for the Lax pairs are given in Sec. \ref{lax}. The
characteristic polynomials, which gives the complete sets of involutive
constant motions for these systems, will also be given there. Sec. \ref{non}%
, is devoted to derive the nonrelativistic limits of these systems which
coincide with the forms given in Refs. \cite{op} and \cite{bcs}. The last
section is brief summary and some discussions.

\sect{$A_{N-1}$-type Ruijsenaars-Schneider model}
\label{ars} \vspace{1pt}
As a relativistic-invariant generalization of the $A_{N-1}$-type
nonrelativistic Calogero-Moser model, the $A_{N-1}$-type
Ruijsenaars-Schneider systems are completely integrable whose integrability
are first showed by Ruijsenaars\cite{r1,r2}. The Lax pairs for this model
have been constructed in Refs. \cite{r1, nksr, bc, kz, s1, s2}. Recent progress
have showed that the compactification of higher dimension SUSY Yang-Mills
theory and Seiberg-Witten theory can be described by this model\cite{bmmm}.
Instanton correction of prepotential associated with $sl_{2}$ $RS$ system have
been calculated in Ref. \cite{ohta}.

\subsection{The Lax operator for A$_{N-1}$ $RS$ model}

Let us briefly give the basics of this model. In terms of the canonical
variables $p_{i}$, $x_{i}(i,j=1,\ldots ,N)$ \ enjoying in the canonical
Poisson bracket

\begin{equation}
\{p_{i},p_{j}\}=\{x_{i},x_{j}\}=0,\mbox{$ \ \ \ \ \ \ \ \ \ \ \ \ \ \ $}%
\{x_{i},p_{j}\}=\delta _{ij},  \label{poisson}
\end{equation}
we give firstly the Hamiltonian of $A_{N-1}$ $RS$ system

\begin{equation}
H_{A_{N-1}}=\sum_{i=1}^{N}\left(e^{p_{i}}\,\prod_{k\neq
i}f(x_{i}-x_{k})+e^{-p_{i}}\,\prod_{k\neq i}g(x_{i}-x_{k})\right).
\label{anhami}
\end{equation}
Notice that in Ref. \cite{r1} Ruijsenaars used another ``gauge'' of the momenta
such that two are connected by the following canonical transformation:
\begin{equation}
x_{i}\longrightarrow x_{i},\ \ \ \ p_{i}\longrightarrow p_{i}+\frac{1}{2}%
ln\prod_{j\neq i}^{N}\frac{f(x_{ij})}{g(x_{ij})}.
\end{equation}

\vspace{1pt}The Lax operator for this model has the form(for the
trigonometric case)
\[
L_{A_{N-1}}=\sum_{i,j=1}^{N}\frac{\sin \gamma }{\sin (x_{i}-x_{j}+\gamma )}\;%
\mathrm{\exp }(p_{j})\ b_{j}E_{ij},
\]
and for the rational case

\begin{equation}
L_{A_{N-1}}=\sum_{i,j=1}^{N}\frac{\gamma }{x_{i}-x_{j}+\gamma }\mathrm{\exp }%
(p_{j})\ b_{j}E_{ij}.
\end{equation}
where
\begin{eqnarray}
b_{j} &:&=\prod_{k\neq j}f(x_{j}-x_{k}),\ \ \ \ \ \ b_{j}^{^{\prime
}}:=\prod_{k\neq j}g(x_{j}-x_{k}),\ \ \ \ \ \ \ \ (E_{ij})_{kl}=\delta
_{ik}\delta _{jl},  \nonumber \\
f(x) &:&=\left\{
\begin{tabular}{ll}
$\frac{\sin (x-\gamma )}{\sin (x)},$ & trigonometric case, \\
$\frac{x-\gamma }{x},$ & rational case,
\end{tabular}
\right.	 \nonumber \\
g(x) &:&=f(x)|_{\gamma \rightarrow -\gamma },\ \ \ \ \ \ \ \ \
x_{ik}:=x_{i}-x_{k},  \label{fg}
\end{eqnarray}
and $\gamma $ denotes the coupling constant.

\vspace{1pt}It is shown in Ref. \cite{s1} that the Lax operator satisfies the
quadratic fundamental Poisson bracket
\begin{equation}
\{L_{1},L_{2}\}=L_{1}\,L_{2}\,a_{1}-a_{2}\,L_{1}\,L_{2}+L_{2}\,s_{1}%
\,L_{1}-L_{1}\,s_{2}\,L_{2},  \label{quad}
\end{equation}
where \ $L_{1}=L_{A_{N-1}}\otimes \emph{1,}L_{2}=\emph{1}\otimes L_{A_{N-1}}%
\emph{\ \ }$and the four matrices read as
\begin{eqnarray}
a_{1} &=&a+w,\quad s_{1}=s-w,  \nonumber \\
a_{2} &=&a+s-s^{\ast }-w,\quad s_{2}=s^{\ast }+w.
\end{eqnarray}
The forms of $a,s,w$ are
\begin{eqnarray}
a &=&\sum_{k\neq j}\mathrm{cot}(x_{k}-x_{j})E_{jk}\otimes E_{kj},  \nonumber
\\
s &=&-\sum_{k\neq j}\frac{1}{\mathrm{sin}(x_{k}-x_{j})}E_{jk}\otimes E_{kk},
\nonumber \\
w &=&\sum_{k\neq j}\mathrm{cot}(x_{k}-x_{j})E_{kk}\otimes E_{jj},
\end{eqnarray}
for the trigonometric case and

\begin{eqnarray}
a &=&\sum_{k\neq j}\frac{1}{x_{k}-x_{j}}E_{jk}\otimes E_{kj},  \nonumber \\
s &=&-\sum_{k\neq j}\frac{1}{x_{k}-x_{j}}E_{jk}\otimes E_{kk},	\nonumber \\
w &=&\sum_{k\neq j}\frac{1}{x_{k}-x_{j}}E_{kk}\otimes E_{jj},
\end{eqnarray}
for the rational case. The $\ast $ symbol means $r^{\ast }=\Pi r\Pi \;\;$with$%
\;\;\Pi =\sum_{k,j=1}^{N}E_{kj}\otimes E_{jk}.$

Noticing that

\begin{equation}
(L_{A_{N-1}}^{-1})_{ij}=\left\{
\begin{tabular}{ll}
$\sum_{i,j=1}^{N}\frac{-\sin \gamma }{\sin (x_{i}-x_{j}-\gamma )}\;\mathrm{%
\exp }(-p_{i})\ b_{j}^{^{\prime }}E_{ij},$ & for trigonometric case, \\
$\sum_{i,j=1}^{N}\frac{-\gamma }{x_{i}-x_{j}-\gamma }\;\mathrm{\exp }%
(-p_{i})\ b_{j}^{^{\prime }}E_{ij},$ & for rational case,
\end{tabular}
\right.
\end{equation}
one can get the characteristic polynomials of $L_{A_{N-1}}$ and
$L_{A_{N-1}}^{-1}$\cite{r3}
\begin{eqnarray}
\det (L_{A_{N-1}}-v\cdot Id)
&=&\sum_{j=0}^{N}(-v)^{n-j}(H_{j}^{+})_{A_{N-1}}, \\
\det (L_{A_{N-1}}^{-1}-v\cdot Id)
&=&\sum_{j=0}^{N}(-v)^{n-j}(H_{j}^{-})_{A_{N-1}},
\end{eqnarray}
where$(H_{0}^{\pm })_{A_{N-1}}=(H_{N}^{\pm })_{A_{N-1}}=1$ and
\begin{eqnarray}
(H_{i}^{+})_{A_{N-1}} &=&\sum_{{{{{\QATOP{J\subset \{1,\ldots ,N\} }{\left|
J\right| =i}}}}}}\exp \left( \sum_{j\in J}p_{j}\right) \,\prod_{{{{{\QATOP{%
j\in J }{k\in \{1,\ldots ,N\}\setminus J}}}}}}f(x_{j}-x_{k}), \\
(H_{i}^{-})_{A_{N-1}} &=&\sum_{{{{{\QATOP{J\subset \{1,\ldots ,N\} }{\left|
J\right| =i}}}}}}\exp \left( \sum_{j\in J}-p_{j}\right) \,\prod_{{{{{\QATOP{%
j\in J }{k\in \{1,\ldots ,N\}\setminus J}}}}}}g(x_{j}-x_{k}).
\end{eqnarray}

\vspace{1pt}Define
\begin{equation}
(H_{i})_{A_{N-1}}=(H_{i}^{+})_{A_{N-1}}+(H_{i}^{-})_{A_{N-1}},
\label{hamiset}
\end{equation}
\vspace{1pt}from the fundamental Poisson bracket Eq.(\ref{quad}), we can
verify that
\begin{equation}
\{(H_{i})_{A_{N-1}},(H_{j})_{A_{N-1}}\}=\{(H_{i}^{\varepsilon
})_{A_{N-1}},(H_{j}^{\varepsilon ^{^{\prime }}})_{A_{N-1}}\}=0,\ \ \ \ \ \
\varepsilon ,\varepsilon ^{^{\prime }}=\pm ,\ \ \ \ i,j=1,\ldots
,N.
\label{aninv}
\end{equation}
In particular, the Hamiltonian Eq.(\ref{anhami}) can be rewritten as
\begin{equation}
H_{A_{N-1}}=(H_{1}^{+})_{A_{N-1}}+(H_{1}^{-})_{A_{N-1}}=%
\sum_{j=1}^{N}(e^{p_{j}}b_{j}+e^{-p_{j}}b_{j}^{^{\prime
}})=Tr(L_{A_{N-1}}+L_{A_{N-1}}^{-1}).
\end{equation}
It should be remarked the set of integrals of motion Eq.(\ref{hamiset}) have
a reflection symmetry which is the key property for the later reduction to $%
C_{n}$ and $BC_{n}$ cases. i.e. if we set

\begin{equation}
p_{i}\longleftrightarrow -p_{i},\mbox{$ \ \ \ \ \ $}x_{i}\longleftrightarrow
-x_{i},	 \label{sym}
\end{equation}
then the Hamiltonians flows $(H_{i})_{A_{N-1}}$ are invariant with respect
to this symmetry.

\subsection{The construction of Lax pair for the $A_{N-1}$ $RS$ model}

As for the $A_{N-1}$ $RS$ model, a generalized Lax pair has been given in Refs.
\cite{r1, nksr, bc, kz, s1, s2}. But there is a common character that the
time-evolution of the Lax matrix $L_{A_{N-1}}$ is associated with the
Hamiltonian $H_{+}$. We will see in the next section that the Lax pair can't
reduce from that \ kind of forms directly. Instead, we give a new Lax pair
which the evolution of $L_{A_{N-1}}$ are associated with the Hamiltonian $\
H_{A_{N-1}}$

\begin{equation}
\dot{L}_{A_{N-1}}=\{L_{A_{N-1}},H_{A_{N-1}}\}=\lbrack
M_{A_{N-1}},L_{A_{N-1}}\rbrack ,  \label{laxeq}
\end{equation}
where $M_{A_{N-1}}$ can be constructed with the help of $(r,s)$ matrices as
follows
\begin{equation}
M_{A_{N-1}}=Tr_{2}((s_{1}-a_{2})(1\otimes (L_{A_{N-1}}-L_{A_{N-1}}^{-1}))).
\end{equation}
The explicit expression of  entries for	 $M_{A_{N-1}}$ is
\begin{eqnarray}
(M_{A_{N-1}})_{ij} &=&\frac{\sin \gamma \cot (x_{ij})}{\sin (x_{ij}+\gamma )}%
e^{p_{j}}b_{j}+\frac{\sin \gamma \cot x_{ij}}{\sin (x_{ij}-\gamma )}%
e^{-p_{i}}b_{j}^{^{\prime }},\ \ \ \ \ \ \ \ \ i\neq j,	 \nonumber \\
(M_{A_{N-1}})_{ii} &=&-\sum_{l\neq i}(\frac{\sin \gamma }{\sin (x_{il})\sin
(x_{il}+\gamma )}e^{p_{l}}b_{l}+\frac{\sin \gamma }{\sin (x_{il})\sin
(x_{il}-\gamma )}e^{-p_{i}}b_{l}^{^{\prime }}),
\end{eqnarray}
for trigonometric case and
\begin{eqnarray}
(M_{A_{N-1}})_{ij} &=&\frac{\gamma }{x_{ij}(x_{ij}+\gamma )}e^{p_{j}}b_{j}+%
\frac{\gamma }{x_{ij}(x_{ij}-\gamma )}e^{-p_{i}}b_{j}^{^{\prime }},\ \ \ \ \
\ \ \ \ i\neq j,  \nonumber \\
(M_{A_{N-1}})_{ii} &=&-\sum_{l\neq i}(\frac{\gamma }{x_{il}(x_{il}+\gamma )}%
e^{p_{l}}b_{l}+\frac{\gamma }{x_{il}(x_{il}-\gamma )}e^{-p_{i}}b_{l}^{^{%
\prime }}),
\end{eqnarray}
for rational case.

\sect{\ Hamiltonian reductions of $C_{n}$ and $BC_{n}$ $RS$ models from $%
A_{N-1} $-type ones}
\label{Ham}
Let us first mention some results about the integrability of Hamiltonian (%
\ref{anhami}). In Ref. \cite{r2} Ruijsenaars demonstrated that the symplectic
structure of $C_{n}$ and \ $BC_{n}$ type $RS$ systems can be proved integrable
by embedding their phase space to a submanifold of $A_{2n-1}$ and $A_{2n}$
type $RS$ ones respectively, while in Refs. \cite{di,di1} and \cite{ko2}, Diejen and
Komori, respectively, gave a series of commuting difference operators which
led to their quantum integrability. However, there are not any results about
their Lax representations so far, i.e. the explicit forms of the Lax
matrixes $L$, associated with a $M$ (respectively) which ensure their Lax
integrability, haven't been proposed up to now except for the special case
of $C_{2}$\cite{kai3}. In this section, we concentrate our treatment to the
exhibition of the explicit forms for general $C_{n}$ and $BC_{n}$ $RS$
systems. Therefore, some previous results, as well as new results, could now
be obtained in a more straightforward manner by using the Lax pairs.

For the convenience of analysis of symmetry, let us first give vector
representation of $A_{N-1}$ Lie algebra. Introducing an $N$ dimensional
orthonormal basis of ${\mathbb R}^{N}$
\begin{equation}
e_{j}\cdot e_{k}=\delta _{j,k},\quad j,k=1,\ldots ,N.
\end{equation}
Then the sets of roots and vector weights are:
\begin{eqnarray}
\Delta &=&\{e_{j}-e_{k}:\quad j,k=1,\ldots ,N\},\quad  \label{anroot} \\
\Lambda &=&\{e_{j}:\quad j=1,\ldots ,N\}.  \label{anwei}
\end{eqnarray}

The dynamical variables are canonical coordinates $\{x_{j}\}$ and their
canonical conjugate momenta $\{p_{j}\}$ with the Poisson brackets of Eq.(\ref
{poisson}) . In general sense, we denote them by $N$ dimensional vectors $x$
and $p$,
\[
x=(x_{1},\ldots ,x_{N})\in {\mathbb R}^{N},\quad p=(p_{1},\ldots ,p_{N})\in {%
\mathbb R}^{N},\quad
\]
so that the scalar products of $x$ and $p$ with the roots $\alpha \cdot x$, $%
p\cdot \beta $, etc. can be defined. The Hamiltonian Eq.(\ref{anhami}) \ can
be rewritten as
\begin{equation}
H_{A_{N-1}}=\sum_{\mu \in \Lambda }\left( \exp \left( \mu \cdot p\right)
\,\prod_{\Delta \ni \beta =\mu -\nu }f(\beta \cdot x){+}\exp \left( -\mu
\cdot p\right) \,\prod_{\Delta \ni \beta =-\mu +\nu }g(\beta \cdot x)\right)
,
\end{equation}
in which $f(x)$ and $g(x)$ is given in Eq.(\ref{fg}) for various choices of
potentials. Here, the condition $\Delta \ni \beta =\mu -\nu $ means that the
summation is over roots $\beta $ such that for $\exists \nu \in \Lambda $

\[
\mu -\nu =\beta \in \Delta .
\]
So does for $\Delta \ni \beta =-\mu +\nu .$

\subsection{$C_{n}$ model}

The set of $C_{n}$ roots consists of two parts, long roots and short roots:
\begin{equation}
\Delta _{C_{n}}=\Delta _{L}\cup \Delta _{S},  \label{cnroot}
\end{equation}
in which the roots are conveniently expressed in terms of an orthonormal
basis of ${\mathbb R}^{n}$:
\begin{eqnarray}
\Delta _{L} &=&\{\pm 2e_{j}:\quad \qquad \ \ j=1,\ldots ,n\},  \nonumber \\
\Delta _{S} &=&\{\pm e_{j}\pm e_{k},:\quad j,k=1,\ldots ,n\}.
\end{eqnarray}
In the vector representation, vector weights $\Lambda $ are

\begin{equation}
\Lambda _{C_{n}}=\{e_{j},-e_{j}:\quad j=1,\ldots ,n\}.
\end{equation}
The Hamiltonian of $C_{n}$ model is given by
\begin{equation}
H_{C_{n}}=\frac{1}{2}\sum_{\mu \in \Lambda _{C_{n}}}\left( \exp \left( \mu
\cdot p\right) \,\prod_{\Delta _{C_{n}}\ni \beta =\mu -\nu }f(\beta \cdot x){%
+}\exp \left( -\mu \cdot p\right) \,\prod_{\Delta _{C_{n}}\ni \beta =-\mu
+\nu }g(\beta \cdot x)\right) .	 \label{cnhami}
\end{equation}
From the above data, we notice that either for $A_{N-1}$ or $C_{n}$ Lie
algebra, any root $\alpha \in \Delta $ can be constructed in terms with
vector weights as $\alpha =\mu -\nu $ where $\mu ,\nu \in \Lambda .$ By
simple comparison of representation between $A_{N-1}$ or $C_{n}$, one can
found that if replacing $e_{j+n}$ with $-e_{j}$ in the vector weights of $%
A_{2n-1}$ algebra, we can obtain the vector weights of $C_{n}$ one. Also
does for the corresponding roots. This hints us it is possible to get the $%
C_{n}$ model by this kind of reduction.

\vspace{1pt}For $A_{2n-1}$ model let us set restrictions on the vector
weights with

\begin{eqnarray}
e_{j+n}+e_{j}=0,\mbox{\ \ \  for \ \ \ }j=1,\ldots ,n,
\end{eqnarray}
which correspond to the following constraints on the phase space of
$A_{2n-1}$-type $RS$ model with
\begin{eqnarray}
G_{i} &\equiv &(e_{i+n}+e_{i})\cdot x=x_{i}+x_{i+n}=0,	\nonumber \\
G_{i+n} &\equiv &(e_{i+n}+e_{i})\cdot p=p_{i}+p_{i+n}=0,\ \ i=1,\ldots
,n,
\label{cncon}
\end{eqnarray}

\noindent Following Dirac's method\cite{Dirac}, we can show

\begin{equation}
\{G_{i},H_{A_{2n-1}}\}\simeq 0,\mbox{\ \ \ for \ \ \ \ }\forall i\in
\{1,\ldots ,2n\},  \label{cnfirst}
\end{equation}
i.e. $H_{A_{2n-1}}$ is the first class Hamiltonian corresponding to the
above constraints Eq. (\ref{cncon}). Here the symbol $\simeq $ represents
that, \ only after calculating the result of left side of the identity,
could we use the conditions of constraints. It should be pointed out that
the most necessary condition ensuring the Eq. (\ref{cnfirst}) is the
symmetry property Eq. (\ref{sym}) for the Hamiltonian Eq. (\ref{anhami}). So
that for arbitrary dynamical variable $A,$ we have

\begin{eqnarray}
\dot{A} &=&\{A,H_{A_{2n-1}}\}_{D}=\{A,H_{A_{2n-1}}\}-\{A,G_{i}\}\Delta
_{ij}^{-1}\{G_{j},H_{A_{2n-1}}\}  \nonumber \\
&\simeq &\{A,H_{A_{2n-1}}\},\qquad \ \ \ \ \ \ \ \ \ \ \ \ \ \ \ \ \ \ \ \ \
\ \ \ \ \ \ \ \ \ \ \ \ \ \ i,j=1,\ldots ,2n,  \label{cnevo}
\end{eqnarray}
where

\vspace{1pt}
\begin{equation}
\Delta _{ij}=\{G_{i},G_{j}\}=2\left(
\begin{array}{cc}
0 & Id \\
-Id & 0
\end{array}
\right) ,
\end{equation}
and the $\{,\}_{D}$ denote the Dirac bracket. By straightforward
calculation, we have the nonzero Dirac brackets of
\begin{eqnarray}
\{x_{i},p_{j}\}_{D} &=&\{x_{i+n},p_{j+n}\}_{D}=\frac{1}{2}\delta _{i,j},
\nonumber \\
\{x_{i},p_{j+n}\}_{D} &=&\{x_{i+n},p_{j}\}_{D}=-\frac{1}{2}\delta _{i,j}.
\label{DBra}
\end{eqnarray}
Using the above data together with the fact that $H_{A_{N-1}}$ is the first
class Hamiltonian (see Eq. (\ref{cnfirst}), we can directly obtain Lax
representation of $\ C_{n}$ $RS$ model by imposing constraints $G_{k}$ on Eq. (%
\ref{laxeq})

\begin{eqnarray}
\{L_{A_{2n-1}},H_{A_{2n-1}}\}_{D}
&=&\{L_{A_{2n-1}},H_{A_{2n-1}}\}|_{G_{k},k=1,...,2n},  \nonumber \\
&=&\lbrack M_{A_{2n-1}},L_{A_{2n-1}}\rbrack |_{G_{k},k=1,...,2n}=\lbrack
M_{C_{n}},L_{C_{n}}\rbrack, \\
\{L_{A_{2n-1}},H_{A_{2n-1}}\}_{D} &=&\{L_{C_{n}},H_{C_{n}}\},  \label{dicn}
\end{eqnarray}
where
\begin{eqnarray}
H_{C_{n}} &=&\frac{1}{2}H_{A_{2n-1}}|_{G_{k},k=1,...,2n},  \nonumber \\
L_{C_{n}} &=&L_{A_{2n-1}}|_{G_{k},k=1,...,2n},	\nonumber \\
M_{C_{n}} &=&M_{A_{2n-1}}|_{G_{k},k=1,...,2n},	\label{reduce}
\end{eqnarray}
so that

\begin{equation}
\dot{L}_{C_{n}}=\{L_{C_{n}},H_{C_{n}}\}=\lbrack M_{C_{n}},L_{C_{n}}\rbrack .
\label{cnLax}
\end{equation}

Nevertheless, the $H_{+}$ is not the first class Hamiltonian, so the Lax
pair given by many authors previously can't reduce to \ $C_{n}$ case
directly by this way.

\subsection{$BC_{n}$ model}

The $BC_{n}$ root system consists of three parts, long, middle and short
roots:
\begin{equation}
\Delta _{BC_{n}}=\Delta _{L}\cup \Delta \cup \Delta _{S},  \label{bcnroot}
\end{equation}
in which the roots are conveniently expressed in terms of an orthonormal
basis of ${\mathbb R}^{n}$:
\begin{eqnarray}
\Delta _{L} &=&\{\pm 2e_{j}:\qquad \quad \ \ j=1,\ldots ,n\},  \nonumber \\
\Delta &=&\{\pm e_{j}\pm e_{k}:\quad j,k=1,\ldots ,n\},	 \nonumber \\
\Delta _{S} &=&\{\pm e_{j}:\qquad \quad \ \ \ \ j=1,\ldots ,n\}.
\label{bcnroots}
\end{eqnarray}
In the vector representation, vector weights $\Lambda $ can be

\begin{equation}
\Lambda _{BC_{n}}=\{e_{j},-e_{j},0:\quad j=1,\ldots ,n\}.
\end{equation}
The Hamiltonian of $BC_{n}$ model is given by
\begin{equation}
H_{BC_{n}}=\frac{1}{2}\sum_{\mu \in \Lambda _{BC_{n}}}\left( \exp \left( \mu
\cdot p\right) \,\prod_{\Delta _{BC_{n}}\ni \beta =\mu -\nu }f(\beta \cdot x)%
{+}\exp \left( -\mu \cdot p\right) \,\prod_{\Delta _{BC_{n}}\ni \beta =-\mu
+\nu }g(\beta \cdot x)\right) .	 \label{bcnhami}
\end{equation}
By similar comparison of representations between $A_{N-1}$ or $BC_{n}$, one
can found that if replacing $e_{j+n}$ with $-e_{j}$ and $e_{2n+1}$with $0$
in the vector weights of $A_{2n}$ Lie algebra, we can obtain the vector
weights of $\allowbreak BC_{n}$ one. Also does for the corresponding roots.
So by the same procedure as $C_{n}$ model, it is expected to get the Lax
representation of $BC_{n}$ model.

\vspace{1pt}For $A_{2n}$ model, we set restrictions on the vector weights
with

\begin{eqnarray}
e_{j+n}+e_{j} &=&0,\mbox{\ \ \ for \ \ \ }j=1,\ldots ,n,  \nonumber \\
e_{2n+1} &=&0,
\end{eqnarray}
which correspond to the following constraints on the phase space of
$A_{2n}$-type $RS$ model with

\begin{eqnarray}
G_{i}^{^{\prime }} &\equiv &(e_{i+n}+e_{i})\cdot x=x_{i}+x_{i+n}=0,
\nonumber \\
G_{i+n}^{^{\prime }} &\equiv &(e_{i+n}+e_{i})\cdot p=p_{i}+p_{i+n}=0,\ \
i=1,\ldots ,n,	\nonumber \\
G_{2n+1}^{^{\prime }} &\equiv &e_{2n+1}\cdot x=x_{2n+1}=0,  \nonumber \\
G_{2n+2}^{^{\prime }} &\equiv &e_{2n+1}\cdot p=p_{2n+1}=0.  \label{bcncon}
\end{eqnarray}

\noindent Similarly, we can show

\begin{equation}
\{G_{i},H_{A_{2n}}\}\simeq 0,\mbox{\ \ \ for \ \ \ \ }\forall i\in
\{1,\ldots ,2n+1,2n+2\}.
\end{equation}
i.e. $H_{A_{2n}}$ is the first class Hamiltonian corresponding to the above
constraints Eq. (\ref{bcncon}). So $L_{BC_{n}}$ and $M_{BC_{n}}$ can be
constructed as follows

\begin{eqnarray}
L_{BC_{n}} &=&L_{A_{2n}}|_{G_{k}^{^{\prime }},k=1,...,2n+2},  \nonumber \\
M_{BC_{n}} &=&M_{A_{2n}}|_{G_{k}^{^{\prime }},k=1,...,2n+2},
\end{eqnarray}
while \ $H_{BC_{n}}$ is

\begin{equation}
H_{BC_{n}}=\frac{1}{2}H_{A_{2n}}|_{G_{k},k=1,...,2n+2},
\end{equation}
due to the similar derivation of Eq.(\ref{cnevo}-\ref{cnLax}).

\sect{ Lax representations of $C_{n} $ and $BC_{n}$ $RS$ models}
\label{lax}

\subsection{$\protect\vspace{1pt}C_{n}$ model}

The Hamiltonian of $C_{n}$ $RS$ system is Eq.(\ref{cnhami}), so the canonical
equations of motion are

\begin{eqnarray}
\dot{x_{i}} &=&\{x_{i},H\}=e^{p_{i}}b_{i}-e^{-p_{i}}b_{i}^{^{\prime }},
\label{mocn1} \\
\dot{p_{i}} &=&\{p_{i},H\}=\sum_{j\neq i}^{n}\{e^{p_{j}}b_{j}(\frac{%
f^{^{\prime }}(x_{ji})}{f(x_{ji})}-\frac{f^{^{\prime }}(x_{j}+x_{i})}{%
f(x_{j}+x_{i})})  \nonumber \\
&&+e^{-p_{j}}b_{j}^{^{\prime }}(\frac{g^{^{\prime }}(x_{ji})}{g(x_{ji})}-%
\frac{g^{^{\prime }}(x_{j}+x_{i})}{g(x_{j}+x_{i})})\}  \nonumber \\
&&-e^{p_{i}}b_{i}(2\frac{f^{^{\prime }}(2x_{i})}{f(2x_{i})}+\sum_{j\neq
i}^{n}(\frac{f^{^{\prime }}(x_{ij})}{f(x_{ij})}+\frac{f^{^{\prime
}}(x_{i}+x_{j})}{f(x_{i}+x_{j})}))  \nonumber \\
&&-e^{-p_{i}}b_{i}^{^{\prime }}(2\frac{g^{^{\prime }}(2x_{i})}{g(2x_{i})}%
+\sum_{j\neq i}^{n}(\frac{g^{^{\prime }}(x_{ij})}{g(x_{ij})}+\frac{%
g^{^{\prime }}(x_{i}+x_{j})}{g(x_{i}+x_{j})})),	 \label{mocn2}
\end{eqnarray}
where

\begin{eqnarray}
f^{^{\prime }}(x) &:&=\frac{df(x)}{dx},\ \ \ \ g^{^{\prime }}(x):=\frac{dg(x)%
}{dx},	\nonumber \\
b_{i} &=&f(2x_{i})\,\prod_{k\neq i}^{n}(f(x_{i}-x_{k})f(x_{i}+x_{k})),
\nonumber \\
b_{i}^{^{\prime }} &=&g(2x_{i})\,\prod_{k\neq
i}^{n}(g(x_{i}-x_{k})g(x_{i}+x_{k})).
\end{eqnarray}

\vspace{1pt}The Lax matrix for $C_{n}$ $RS$ model can be written
in the following  form for the rational case
\begin{equation}
(L_{C_{n}})_{\mu \nu }=e^{\nu \cdot p}b_{\nu }\frac{\gamma }{(\mu -\nu
)\cdot x+\gamma },  \label{Lmatrix}
\end{equation}
which is a $2n\times 2n$ matrix whose indices are labelled by the vector
weights, denoted by \ $\mu ,\nu \in \Lambda _{C_{n}}$, $M_{C_{n}}$
can be written as

\begin{equation}
M_{C_{n}}=D+Y,	\label{Mmatrix}
\end{equation}
where
\begin{eqnarray}
Y_{\mu \nu } &=&e^{\nu \cdot p}b_{\nu }\frac{\gamma }{((\mu -\nu )\cdot
x)((\mu -\nu )\cdot x+\gamma )}	 \nonumber \\
&&+e^{-\mu \cdot p}b_{\nu }^{^{\prime }}\frac{\gamma }{((\mu -\nu )\cdot
x)((\mu -\nu )\cdot x-\gamma )},  \label{ratiY} \\
D_{\mu \mu } &=&-\sum_{\nu \neq \mu }(e^{\nu \cdot p}b_{\nu }\frac{\gamma }{%
((\mu -\nu )\cdot x)((\mu -\nu )\cdot x+\gamma )}  \nonumber \\
&&+e^{-\mu \cdot p}b_{\nu }^{^{\prime }}\frac{\gamma }{((\mu -\nu )\cdot
x)((\mu -\nu )\cdot x-\gamma )})  \nonumber \\
&=&-\sum_{\nu \neq \mu }Y_{\mu \nu },  \label{ratiD}
\end{eqnarray}
and
\begin{eqnarray}
b_{\mu } &=&\prod_{\Delta _{C_{n}}\ni \beta =\mu -\nu }f(\beta \cdot x),
\nonumber \\
b_{\mu }^{^{\prime }} &=&\prod_{\Delta _{C_{n}}\ni \beta =\mu -\nu }g(\beta
\cdot x).  \label{cnaux}
\end{eqnarray}
For the trigonometric case, we have

\begin{equation}
(L_{C_{n}})_{\mu \nu }=e^{\nu \cdot p}b_{\nu }\frac{\sin \gamma }{\sin ((\mu
-\nu )\cdot x+\gamma )},  \label{trigL}
\end{equation}
and

\begin{equation}
M_{C_{n}}=D+Y,	\label{trigM}
\end{equation}
where
\begin{eqnarray}
Y_{\mu \nu } &=&e^{\nu \cdot p}b_{\nu }\frac{\sin \gamma \cot ((\mu -\nu
)\cdot x)}{\sin ((\mu -\nu )\cdot x+\gamma )}+e^{-\mu \cdot p}b_{\nu
}^{^{\prime }}\frac{\sin \gamma \cot ((\mu -\nu )\cdot x)}{\sin ((\mu -\nu
)\cdot x-\gamma )},  \label{trigY} \\
D_{\mu \mu } &=&-\sum_{\nu \neq \mu }(e^{\nu \cdot p}b_{\nu }\frac{\sin
\gamma }{\sin ((\mu -\nu )\cdot x)\sin ((\mu -\nu )\cdot x+\gamma )}
\nonumber \\
&&+e^{-\mu \cdot p}b_{\nu }^{^{\prime }}\frac{\sin \gamma }{\sin ((\mu -\nu
)\cdot x)\sin ((\mu -\nu )\cdot x-\gamma )})  \nonumber \\
&=&-\sum_{\nu \neq \mu }\frac{Y_{\mu \nu }}{\cos ((\mu -\nu )\cdot x)},
\label{trigD}
\end{eqnarray}
where $b_{\mu },b_{\mu }^{^{\prime }}$ take the value as Eq.( \ref{cnaux})
with the trigonometric forms of $f(x)$ and $g(x)$.

The $L_{C_{n}},M_{C_{n}}$ satisfies the Lax equation

\begin{equation}
\dot{L}_{C_{n}}=\{L_{C_{n}},H_{C_{n}}\}=\lbrack M_{C_{n}},L_{C_{n}}\rbrack ,
\end{equation}
which equivalent to the equations of motion Eq.(\ref{mocn1}) and Eq.(\ref
{mocn2}). The Hamiltonian $H_{C_{n}}$ can be rewritten
as the trace of $L_{C_{n}}$

\begin{equation}
H_{C_{n}}=trL_{C_{n}}=\frac{1}{2}\sum_{\mu \in \Lambda _{C_{n}}}(e^{\mu
\cdot p}b_{\mu }+e^{-\mu \cdot p}b_{\mu }^{^{\prime }}).
\end{equation}
The characteristic polynomial of the Lax matrix $L_{C_{n}}$ generates the involutive
Hamiltonians

\begin{equation}
\det (L_{C_{n}}-v\cdot
Id)=\sum_{j=0}^{n-1}(-1)^{j}(v^{j}+v^{2n-j})(H_{j})_{C_{n}}+(-v)^{n}%
\;(H_{n})_{C_{n}},
\end{equation}
where $(H_{0})_{C_{n}}=1$, and $\ (H_{i})_{C_{n}}$ Poisson commute

\begin{equation}
\{(H_{i})_{C_{n}},(H_{j})_{C_{n}}\}=0,\mbox{$ \ \ \ \ \ \ \ \
$}i,j=1,...,n.
\end{equation}
This can be deduced by verbose but straightforward calculation to verify
that the $(H_{i})_{A_{2n-1}},i=1,...2n$ is the first class \ Hamiltonian
with respect to the constraints Eq.(\ref{cncon}), using Eq.(\ref{aninv}), (%
\ref{cnevo}) and the first formula of Eq.(\ref{reduce}).

The explicit form of $(H_{l})_{C_{n}}$ are
\begin{eqnarray}
(H_{l})_{C_{n}}&=&\sum_{\stackrel{J\subset \{1,\ldots ,n\},\,|J|\leq l}{%
\varepsilon _{j}=\pm 1,\,j\in J}}\mathrm{\exp }(p_{\varepsilon
J})\,F_{\varepsilon J;\,J^{c}}\,U_{J^{c},\,l-|J|},\;\;\;\;\;\;\;\;l=1,\ldots
,n,
\end{eqnarray}
with

\begin{eqnarray}
p_{\varepsilon J} &=&\sum_{j\in J}\;\varepsilon _{j}\,p_{j},  \nonumber \\
F_{\varepsilon J;\,K} &=&\,\prod_{\stackrel{j,j^{\prime }\in J}{j<j^{\prime }%
}}f^{2}(\varepsilon _{j}x_{j}+\varepsilon _{j^{\prime }}x_{j^{\prime
}})\,\prod_{\stackrel{j\in J}{k\in K}}f(\varepsilon
_{j}x_{j}+x_{k})f(\varepsilon _{j}x_{j}-x_{k})\prod_{j\in J}f(2\varepsilon
_{j}x_{j}),  \nonumber \\
U_{I,p} &=&\sum_{\stackrel{I^{\prime }\subset I}{|I^{\prime }|=\lbrack p/2\rbrack }}\prod_{%
\stackrel{j\in I^{\prime }}{k\in I\backslash I^{\prime }}%
\,}f(x_{jk})f(x_{j}+x_{k})g(x_{jk})g(x_{j}+x_{k})\left\{
\begin{array}{c}
0,\mbox{ \ ($p$ odd),} \\
1,\mbox{ \ ($p$ even).}
\end{array}
\right.
\end{eqnarray}
Here, $[p/2]$ denotes the integer part of $p/2$.
As an example, for $C_{2}$ $RS$ model, the independent Hamiltonian flows $%
(H_{1})_{C_{2}}$ and $(H_{2})_{C_{2}}$ generated by the Lax matrix $%
L_{C_{2}} $ are\cite{kai3}

\begin{eqnarray}
(H_{1})_{C_{2}} &=&H_{C_{2}}=e^{p_{1}}f(2x_{1})\,f(x_{12})f(x_{1}+x_{2})
\nonumber \\
&&+e^{-p_{1}}g(2x_{1})\,g(x_{12})g(x_{1}+x_{2})	 \nonumber \\
&&+e^{p_{2}}f(2x_{2})\,f(x_{21})f(x_{2}+x_{1})	\nonumber \\
&&+e^{-p_{2}}g(2x_{2})\,g(x_{21})g(x_{2}+x_{1}), \\
(H_{2})_{C_{2}} &=&e^{p_{1}+p_{2}}f(2x_{1})\,(f(x_{1}+x_{2}))^{2}f(2x_{2})
\nonumber \\
&&+e^{-p_{1}-p_{2}}g(2x_{1})\,(g(x_{1}+x_{2}))^{2}g(2x_{2})  \nonumber \\
&&+e^{p_{1}-p_{2}}f(2x_{1})\,(f(x_{12}))^{2}f(-2x_{2})	\nonumber \\
&&+e^{p_{2}-p_{1}}g(2x_{1})\,(g(x_{12}))^{2}g(-2x_{2})	\nonumber \\
&&+2f(x_{12})\,g(x_{12})\,f(x_{1}+x_{2})g(x_{1}+x_{2}).
\end{eqnarray}

\subsection{$BC_{n}$ model}

The Hamiltonian $BC_{n}$ model is expressed in Eq.(\ref{bcnhami}), so the
canonical equations of motion are

\begin{eqnarray}
\dot{x_{i}} &=&\{x_{i},H\}=e^{p_{i}}b_{i}-e^{-p_{i}}b_{i}^{^{\prime }},
\label{mobcn1} \\
\dot{p_{i}} &=&\{p_{i},H\}=\sum_{j\neq i}^{n}\{e^{p_{j}}b_{j}(\frac{%
f^{^{\prime }}(x_{ji})}{f(x_{ji})}-\frac{f^{^{\prime }}(x_{j}+x_{i})}{%
f(x_{j}+x_{i})})  \nonumber \\
&&+e^{-p_{j}}b_{j}^{^{\prime }}(\frac{g^{^{\prime }}(x_{ji})}{g(x_{ji})}-%
\frac{g^{^{\prime }}(x_{j}+x_{i})}{g(x_{j}+x_{i})})\}  \nonumber \\
&&-e^{p_{i}}b_{i}(\frac{f^{^{\prime }}(x_{i})}{f(x_{i})}+2\frac{f^{^{\prime
}}(2x_{i})}{f(2x_{i})}+\sum_{j\neq i}^{n}(\frac{f^{^{\prime }}(x_{ij})}{%
f(x_{ij})}+\frac{f^{^{\prime }}(x_{i}+x_{j})}{f(x_{i}+x_{j})}))	 \nonumber \\
&&-e^{-p_{i}}b_{i}^{^{\prime }}(\frac{g^{^{\prime }}(x_{i})}{g(x_{i})}+2%
\frac{g^{^{\prime }}(2x_{i})}{g(2x_{i})}+\sum_{j\neq i}^{n}(\frac{%
g^{^{\prime }}(x_{ij})}{g(x_{ij})}+\frac{g^{^{\prime }}(x_{i}+x_{j})}{%
g(x_{i}+x_{j})}))  \nonumber \\
&&-b_{0}(\frac{f^{^{\prime }}(x_{i})}{f(x_{i})}+\frac{g^{^{\prime }}(x_{i})}{%
g(x_{i})}),  \label{mobcn2}
\end{eqnarray}
where

\begin{eqnarray}
b_{i} &=&f(x_{i})f(2x_{i})\,\prod_{k\neq
i}^{n}(f(x_{i}-x_{k})f(x_{i}+x_{k})),  \nonumber \\
b_{i}^{^{\prime }} &=&g(x_{i})g(2x_{i})\,\prod_{k\neq
i}^{n}(g(x_{i}-x_{k})g(x_{i}+x_{k})),  \nonumber \\
b_{0} &=&\prod_{i=1}^{n}(f(x_{i})g(x_{i}).
\end{eqnarray}

The Lax pair for $BC_{n}$ $RS$ model can be constructed as the form of Eq.(\ref
{Lmatrix})-(\ref{trigD}) where one should replace the matrices labels with $%
\mu ,\nu \in \Lambda _{BC_{n}},$ and roots with $\beta \in \Delta _{BC_{n}}.$

The Hamiltonian $H_{BC_{n}}$ can be rewritten as the trace
of $L_{BC_{n}}$

\begin{equation}
H_{BC_{n}}=trL_{BC_{n}}=\frac{1}{2}\sum_{\mu \in \Lambda _{BC_{n}}}(e^{\mu
\cdot p}b_{\mu }+e^{-\mu \cdot p}b_{\mu }^{^{\prime }}).
\end{equation}
The characteristic polynomial of the Lax matrix $L$ generates the involutive
Hamiltonians

\begin{equation}
\det (L_{BC_{n}}-v\cdot
Id)=\sum_{j=0}^{n}(-1)^{j}(v^{j}-v^{2n+1-j})(H_{j})_{BC_{n}},
\end{equation}
where $(H_{0})_{BC_{n}}=1$ and \ $(H_{i})_{BC_{n}}$ Poisson commute

\begin{equation}
\{(H_{i})_{BC_{n}},(H_{j})_{BC_{n}}\}=0,\mbox{$ \ \ \ \ \ \ \ \ \ \ \ $}%
i,j=1,...,n.
\end{equation}
This can be deduced similarly to $C_{n}$ case to verify that the $%
(H_{i})_{A_{2n}},i=1,...2n$ is the first class \ Hamiltonian with respect to
the constraints Eq.(\ref{bcncon}).

The explicit form of $(H_{l})_{BC_{n}}$ are

\vspace{1pt}
\begin{eqnarray}
(H_{l})_{BC_{n}}=\sum_{\stackrel{J\subset \{1,\ldots ,n\},\,|J|\leq l}{%
\varepsilon _{j}=\pm 1,\,j\in J}}\mathrm{\exp }(p_{\varepsilon
J})\,F_{\varepsilon J;\,J^{c}}\,U_{J^{c},\,l-|J|},\;\;\;\;\;\;\;\;l=1,\ldots
,n,
\end{eqnarray}
with

\begin{eqnarray}
p_{\varepsilon J} &=&\sum_{j\in J}\;\varepsilon _{j}\,p_{j},  \nonumber \\
F_{\varepsilon J;\,K} &=&\,\prod_{\stackrel{j,j^{\prime }\in J}{j<j^{\prime }%
}}f^{2}(\varepsilon _{j}x_{j}+\varepsilon _{j^{\prime }}x_{j^{\prime
}})\,\prod_{\stackrel{j\in J}{k\in K}}f(\varepsilon
_{j}x_{j}+x_{k})f(\varepsilon _{j}x_{j}-x_{k})\prod_{j\in J}f(2\varepsilon
_{j}x_{j})\prod_{j\in J}f(\varepsilon _{j}x_{j}),  \nonumber \\
U_{I,p} &=&\sum_{\stackrel{I^{\prime }\subset I}{|I^{\prime }|=\lbrack p/2\rbrack }}\prod_{%
\stackrel{j\in I^{\prime }}{k\in I\backslash I^{\prime }}%
\,}f(x_{jk})f(x_{j}+x_{k})g(x_{jk})g(x_{j}+x_{k})\left\{
\begin{array}{c}
\prod_{i\in I\backslash I^{\prime }}f(x_{i})g(x_{i}),\mbox{ \ ($p$ odd),} \\
\prod_{i^{\prime }\in I^{\prime }}f(x_{i^{\prime }})g(x_{i^{\prime }}),%
\mbox{
\ \ ($p$ even).}
\end{array}
\right.	 \nonumber \\
&&
\end{eqnarray}

\sect{Nonrelativistic limit to the Calogero-Moser system}
\label{non}

\subsection{\protect\vspace{1pt}Limit to $C_{n}$ $CM$ model}

The Nonrelativistic limit can be achieved by rescaling \ $p_{i}\longmapsto
\beta p_{i}$, $\gamma \longmapsto \beta \gamma $ while letting $\beta
\longmapsto 0,$ and making a canonical transformation

\begin{eqnarray}
\left\{
\begin{tabular}{ll}
$p_{i}\longmapsto p_{i}+\gamma \{\frac{1}{2x_{i}}+\sum_{k\neq i}^{n}(\frac{1%
}{x_{ik}}+\frac{1}{x_{i}+x_{k}})\},\ $ & rational case, \\
$p_{i}\longmapsto p_{i}+\gamma \{\cot (2x_{i})+\sum_{k\neq i}^{n}(\cot
(x_{ik})+\cot (x_{i}+x_{k}))\},\ \ $ & trigonometric case,
\end{tabular}
\right.
\end{eqnarray}
\noindent such that
\begin{eqnarray}
L &\longmapsto &Id+\beta L_{CM}+O(\beta ^{2}), \\
M &\longmapsto &2\beta M_{CM}+O(\beta ^{2}),
\end{eqnarray}
and

\begin{eqnarray}
H\longmapsto 2n+2\beta ^{2}H_{CM}+O(\beta ^{2}).
\end{eqnarray}
$L_{CM}$ can be expressed as
\begin{eqnarray}
L_{CM}=\left(
\begin{array}{ll}
A_{CM} & B_{CM} \\
-B_{CM} & -A_{CM}
\end{array}
\right),
\end{eqnarray}
where

\begin{eqnarray}
\ (A_{CM})_{ii}&=&p_{i},\ \ \ \ \ \ \ (B_{CM})_{ij}=\frac{\gamma }{%
x_{i}+x_{j}},  \nonumber \\
(A_{CM})_{ij}&=&\frac{\gamma }{x_{ij}}, \ \ \ \ \ (\ i\neq j),
\end{eqnarray}
for the rational case, and

\begin{eqnarray}
(A_{CM})_{ii} &=&p_{i},\ \ \ \ \ \ \ (B_{CM})_{ij}=\frac{\gamma }{\sin
(x_{i}+x_{j})},	 \nonumber \\
(A_{CM})_{ij} &=&\frac{\gamma }{\sin (x_{ij})},\ \ \ \ (i\neq
j),
\end{eqnarray}
for the trigonometric case.

$M_{CM}$ is
\begin{eqnarray}
M_{CM}=\left(
\begin{array}{ll}
\mathcal{A}_{CM} & \mathcal{B}_{CM} \\
\mathcal{B}_{CM} & \mathcal{A}_{CM}
\end{array}
\right),
\end{eqnarray}
as for the rational case

\begin{eqnarray}
(\mathcal{A}_{CM})_{ii} &=&-\sum_{k\neq i}^{n}(\frac{\gamma }{x_{ik}^{2}}+%
\frac{\gamma }{(x_{i}+x_{k})^{2}})-\frac{\gamma }{(2x_{i})^{2}},\ \ \ \ (%
\mathcal{B}_{CM})_{ij}=\frac{\gamma }{(x_{i}+x_{j})^{2}},  \nonumber \\
(\mathcal{A}_{CM})_{ij} &=&\frac{\gamma }{x_{ij}^{2}},\ \ \ \ \ (\ i\neq j),
\end{eqnarray}
which identified with the results of Refs. \cite{op} and \cite{bcs},

\noindent and for the trigonometric case

\begin{eqnarray}
(\mathcal{A}_{CM})_{ii} &=&-\sum_{k\neq i}^{n}(\frac{\gamma }{\sin ^{2}x_{ik}%
}+\frac{\gamma }{\sin ^{2}(x_{i}+x_{k})})-\frac{\gamma }{\sin ^{2}(2x_{i})}%
,\ \ \ \ (\mathcal{B}_{CM})_{ij}=\frac{\gamma \cos (x_{i}+x_{j})}{\sin
^{2}(x_{i}+x_{j})},  \nonumber \\
(\mathcal{A}_{CM})_{ij} &=&\frac{\gamma \cos (x_{ij})}{\sin ^{2}x_{ij}},\ \
\ \ \ (\ i\neq j),
\end{eqnarray}
which coincide with the form given in Ref. \cite{op} up to a diagonalized matrix
together with a suitable choose of coupling parameters.

The Hamiltonian of $C_{n}$-type $CM$ model can be given by

\begin{eqnarray}
H_{CM} &=&\frac{1}{2}\sum_{k=1}^{n}p_{k}^{2}-\gamma ^{2}\sum_{k<i}^{n}(\frac{%
1}{x_{ik}^{2}}+\frac{1}{(x_{i}+x_{k})^{2}})-\frac{\gamma ^{2}}{2}%
\sum_{i=1}^{n}\frac{1}{(2x_{i})^{2}}  \nonumber \\
&=&\frac{1}{4}trL^{2},\ \ \ \ \ \ \ \ \ \ \ \ \ \ \
\mbox{for the rational
case}, \\
H_{CM} &=&\frac{1}{2}\sum_{k=1}^{n}p_{k}^{2}-\gamma ^{2}\sum_{k\neq i}^{n}(%
\frac{1}{\sin ^{2}x_{ik}}+\frac{1}{\sin ^{2}(x_{i}+x_{k})})-\frac{\gamma ^{2}%
}{2}\sum_{i=1}^{n}\frac{1}{\sin ^{2}(2x_{i})}  \nonumber \\
&=&\frac{1}{4}trL^{2},\ \ \ \ \ \ \ \ \ \ \ \ \ \ \
\mbox{for the trigonometric
case}.
\end{eqnarray}
The $L_{CM}$, $M_{CM}$ satisfies the Lax equation

\begin{eqnarray}
\dot{L}_{CM}=\{L_{CM},H_{CM}\}=\lbrack M_{CM},L_{CM}\rbrack .
\end{eqnarray}

\subsection{Limit to $BC_{n}$ $CM$ model}

The Nonrelativistic limit of $BC_{n}$ model can also be achieved by
rescaling \ $p_{i}\longmapsto \beta p_{i}$, $\gamma \longmapsto \beta \gamma
$ while letting $\beta \longmapsto 0,$ and making the following canonical
transformation

\begin{eqnarray}
&&\left\{
\begin{tabular}{ll}
$p_{i}\longmapsto p_{i}+\gamma \{\frac{1}{x_{i}}+\frac{1}{2x_{i}}%
+\sum_{k\neq i}^{n}(\frac{1}{x_{ik}}+\frac{1}{x_{i}+x_{k}})\},\ $ & rational
case, \\
$p_{i}\longmapsto p_{i}+\gamma \{\cot (x_{i})+\cot (2x_{i})+\sum_{k\neq
i}^{n}(\cot (x_{ik})+\cot (x_{i}+x_{k}))\},$ & trigonometric case,
\end{tabular}
\right.	 \nonumber \\
&&
\end{eqnarray}
such that
\begin{eqnarray}
L &\longmapsto &Id+\beta L_{CM}+O(\beta ^{2}), \\
M &\longmapsto &2\beta M_{CM}+O(\beta ^{2}),
\end{eqnarray}
and

\begin{eqnarray}
H\longmapsto (2n+1)+2\beta ^{2}H_{CM}+O(\beta ^{2}).
\end{eqnarray}

$L_{CM}$ can be expressed as
\begin{eqnarray}
L_{CM}=\left(
\begin{array}{lll}
A_{CM} & B_{CM} & E_{CM} \\
-B_{CM} & -A_{CM} & -E_{CM} \\
-(E_{CM})^{t} & (E_{CM})^{t} & G_{CM}
\end{array}
\right),
\end{eqnarray}
where

\begin{eqnarray}
\ (A_{CM})_{ii} &=&p_{i},\ \ \ \ \ \ \ (B_{CM})_{ij}=\frac{\gamma }{%
x_{i}+x_{j}},\ \ \ \ \ \ (E_{CM})_{i1}=\frac{1}{x_{i}},\ \ \ \ \ \ G_{CM}=0,
\nonumber \\
(A_{CM})_{ij} &=&\frac{\gamma }{x_{ij}},\ \ \ \ \ (\ i\neq j),\ \ \ \ \ \
\end{eqnarray}
for the rational case, and

\begin{eqnarray}
(A_{CM})_{ii} &=&p_{i},\ \ \ \ \ \ \ (B_{CM})_{ij}=\frac{\gamma }{\sin
(x_{i}+x_{j})},\ \ \ \ (E_{CM})_{i1}=\frac{1}{\sin x_{i}},\ \ \ \ \ \
G_{CM}=0,  \nonumber \\
(A_{CM})_{ij} &=&\frac{\gamma }{\sin (x_{ij})},\ \ \ \ (i\neq
j),
\end{eqnarray}
for the trigonometric case.

$M_{CM}$ is
\begin{equation}
M_{CM}=\left(
\begin{array}{lll}
\mathcal{A}_{CM} & \mathcal{B}_{CM} & \mathcal{E}_{CM} \\
\mathcal{B}_{CM} & \mathcal{A}_{CM} & \mathcal{E}_{CM} \\
(\mathcal{E}_{CM})^{t} & (\mathcal{E}_{CM})^{t} & \mathcal{G}_{CM}
\end{array}
\right) ,
\end{equation}
where $t$ denotes the transposition. As for the rational case, the forms of $%
\ \mathcal{A,B,E,G}$ are

\begin{eqnarray}
(\mathcal{A}_{CM})_{ii} &=&-\sum_{k\neq i}^{n}(\frac{\gamma }{x_{ik}^{2}}+%
\frac{\gamma }{(x_{i}+x_{k})^{2}})-\frac{\gamma }{(2x_{i})^{2}}-\frac{\gamma
}{(x_{i})^{2}},	 \nonumber \\
\ \ (\mathcal{B}_{CM})_{ij} &=&\frac{\gamma }{(x_{i}+x_{j})^{2}},\ \ \ \ \ \
(\mathcal{E}_{CM})_{i1}=\frac{\gamma }{(x_{i})^{2}},  \nonumber \\
(\mathcal{A}_{CM})_{ij} &=&\frac{\gamma }{x_{ij}^{2}},\ \ \ \ \ (\ i\neq
j),\ \ \ \ \mathcal{G}_{CM}=-\sum_{k=1}^{n}\frac{2\gamma }{x_{k}^{2}},
\end{eqnarray}
which identified with the results of Refs. \cite{op} and \cite{bcs},

\noindent and for the trigonometric case

\begin{eqnarray}
(\mathcal{A}_{CM})_{ii} &=&-\sum_{k\neq i}^{n}(\frac{\gamma }{\sin ^{2}x_{ik}%
}+\frac{\gamma }{\sin ^{2}(x_{i}+x_{k})})-\frac{\gamma }{\sin ^{2}(2x_{i})}-%
\frac{\gamma }{\sin ^{2}(x_{i})},  \nonumber \\
\ (\mathcal{B}_{CM})_{ij} &=&\frac{\gamma \cos (x_{i}+x_{j})}{\sin
^{2}(x_{i}+x_{j})},\ \ \ \ \ \ \ \ (\mathcal{E}_{CM})_{i1}=\frac{\gamma \cos
x_{i}}{\sin ^{2}x_{i}},	 \nonumber \\
(\mathcal{A}_{CM})_{ij} &=&\frac{\gamma \cos (x_{ij})}{\sin ^{2}x_{ij}},\ \
\ \ (\ i\neq j),\ \ \ \mathcal{G}_{CM}=-\sum_{k=1}^{n}\frac{2\gamma }{\sin
^{2}x_{k}},
\end{eqnarray}
which coincide with the form given in Ref. \cite{op} up to a diagonalized matrix
together with a suitable choose of coupling parameters.

The Hamiltonian of $BC_{n}$-type $CM$ model can be given by

\begin{eqnarray}
H_{CM} &=&\frac{1}{2}\sum_{k=1}^{n}p_{k}^{2}-\gamma ^{2}\sum_{k<i}^{n}(\frac{%
1}{x_{ik}^{2}}+\frac{1}{(x_{i}+x_{k})^{2}})-\frac{\gamma ^{2}}{2}%
\sum_{i=1}^{n}(\frac{1}{(2x_{i})^{2}}+\frac{2}{(x_{i})^{2}})  \nonumber \\
&=&\frac{1}{4}trL^{2},\ \ \ \ \ \ \ \ \ \ \ \ \ \ \
\mbox{for the rational
case}, \\
H_{CM} &=&\frac{1}{2}\sum_{k=1}^{n}p_{k}^{2}-\gamma ^{2}\sum_{k<i}^{n}(\frac{%
1}{\sin ^{2}x_{ik}}+\frac{1}{\sin ^{2}(x_{i}+x_{k})})-\frac{\gamma ^{2}}{2}%
\sum_{i=1}^{n}(\frac{1}{\sin ^{2}(2x_{i})}+\frac{2}{\sin ^{2}(x_{i})})
\nonumber \\
&=&\frac{1}{4}trL^{2},\ \ \ \ \ \ \ \ \ \ \ \ \ \ \
\mbox{for the trigonometric
case}.
\end{eqnarray}
The $L_{CM}$, $M_{CM}$ satisfies the Lax equation

\begin{eqnarray}
\dot{L}_{CM}=\{L_{CM},H_{CM}\}=\lbrack M_{CM},L_{CM}\rbrack .
\end{eqnarray}

\sect{Summary and discussions}
\label{sum}
In this paper, we have proposed
the Lax pairs for rational, trigonometric $C_{n}$ and $BC_{n}$ $RS$ models.
Involutive Hamiltonians are showed to be generated by the characteristic
polynomial of the corresponding Lax matrix. In the nonrelativistic limit,
the system leads to $CM$ systems associated with the root systems of $C_{n}$
and $BC_{n}$ which are known previously. There are still many open problems,
for example, it seems to be a challenging subject to carry out the Lax pairs
with as many independent coupling constants as independent Weyl orbits in
the set of roots, as done for the Calogero-Moser systems\cite
{op,bcs,bcs2,bcs3,bcs1,hp1}. What is also interesting may generalize the
results obtained in this paper to the systems associated with all of other
Lie Algebras even to those associated with all the finite reflection groups%
\cite{bcs1} which including models based on the non-crystallographic root
systems and those based on crystallographic root systems.

\section*{Acknowledgement}

One of the authors K. Chen is grateful to professors K. J. Shi and L. Zhao
for their encouragement. This work has been supported financially by the
National Natural Science Foundation of China.

\vspace{1pt}

\vspace{1pt}

\end{document}